\input epsf.tex
\documentclass[12pt]{article}
\usepackage{graphicx}
\csname @addtoreset\endcsname{equation}{section}

\def\bseq{\begin{subequation}}  
\def\eseq{\end{subequation}}
\def\bsea{\begin{subeqnarray}}  
\def\esea{\end{subeqnarray}}

\newcommand{\bbox}{\lower.2ex\hbox{$\Box$}}

\newcommand{\beq}{\begin{equation}}
\newcommand{\eeq}{\end{equation}}
\newcommand{\bea}{\begin{eqnarray}}
\newcommand{\eea}{\end{eqnarray}}
\newcommand{\ena}{\end{eqnarray}}

\renewcommand{\a}{\alpha}
\renewcommand{\b}{\beta}

\renewcommand{\d}{\delta}

\newcommand{\G}{\Gamma}

\newcommand{\e}{\epsilon}
\newcommand{\z}{\zeta}

\renewcommand{\l}{\lambda}

\newcommand{\m}{\mu}

\newcommand{\n}{\nu}

\newcommand{\p}{\pi}

\newcommand{\Db}{\bar{D}}

\newcommand{\Phib}{\bar{\Phi}}

\newcommand{\ad}{{\dot{\alpha}}}

\begin{document}
\begin{titlepage}
\begin{flushright}
Bicocca--FT--99--30 \\
KUL--TF--99/38\\
IFUM--648--FT \\
\end{flushright}
\vspace{.3cm}
\begin{center}
{\Large \bf Two-point functions of chiral operators in ${\cal N}=4$ SYM at
order $g^4$ }
\vfill

{\large \bf Silvia Penati$^1$,
Alberto Santambrogio$^2$ and
Daniela Zanon$^3$}\\
\vfill

{\small
$^1$ Dipartimento di Fisica dell'Universit\`a di Milano--Bicocca
and\\ INFN, Sezione di Milano, Via Celoria 16,
20133 Milano, Italy\\
\vspace*{0.4cm}
$^2$  Instituut voor Theoretische Fysica - Katholieke Universiteit Leuven
\\Celestijnenlaan 200D B--3001 Leuven, Belgium\\
\vspace*{0.4cm}
$^3$ Dipartimento di Fisica dell'Universit\`a di Milano
and\\ INFN, Sezione di Milano, Via Celoria 16,
20133 Milano, Italy\\}
\end{center}
\vfill
\begin{center}
{\bf Abstract}
\end{center}
{\small We compute two-point functions of chiral operators ${\rm Tr} \Phi^3$
in ${\cal N}=4$
$SU(N)$ supersymmetric Yang-Mills theory to the order $g^4$ in perturbation
theory.
We perform explicit calculations using  ${\cal N}=1$ superspace techniques and
find that perturbative corrections to the correlators vanish for all $N$.
While at order $g^2$ the cancellations can be ascribed to
the nonrenormalization theorem valid for correlators of operators in the
same multiplet as
the stress tensor, at order $g^4$ this argument no longer applies and the
actual cancellation occurs in a highly nontrivial way.
Our result
is obtained in complete generality, without the need of additional
conjectures or
assumptions. It gives further support to the belief that such correlators
are not
renormalized to {\em all} orders in $g$ and to {\em all} orders in $N$.}
\vspace{2mm} \vfill \hrule width 3.cm
\begin{flushleft}
e-mail: silvia.penati@mi.infn.it\\
e-mail: alberto.santambrogio@fys.kuleuven.ac.be\\
e-mail: daniela.zanon@mi.infn.it
\end{flushleft}
\end{titlepage}

 \section{Introduction}

Recently much evidence has been provided in testing the conjectured
equivalence of
type $IIB$ superstring theory on anti-de-Sitter space (AdS) times a compact
manifold
to the ${\cal N}=4$ supersymmetric $SU(N)$ Yang-Mills conformal field theory
living
on the boundary, in the large $N$ limit and at large 't Hooft coupling
$\lambda=g^2 N/4\pi$ ($g^2$ being the Yang-Mills coupling constant)
\cite{adscft}.
According to this correspondence correlation functions of
operators in the conformal field theory are mapped
to appropriate on-shell amplitudes of superstring theory in the bulk AdS
background. ${\cal N}=4$ chiral primary operators
${\rm Tr} \Phi^k\equiv {\rm Tr} \{ \Phi^{\{ i_1}(z)\Phi^{i_2}(z)\cdots
\Phi^{i_k\} }(z)\}$, in
the symmetric, traceless representation of $SU(4)$, play a special role in
exploring
non-perturbative statements concerning the above mentioned connection. These
are local operators of the lowest scaling dimension in a given irreducible
representation of the superconformal
algebra $SU(2,2|4)$, and belong to short multiplets which are chiral under
a ${\cal N}=1$ subalgebra.
In the large $N$ limit they correspond to Kaluza Klein modes in the AdS
supergravity sector.
In the special case of $k=2$, two- and three-point correlators
are given by their free-field theory values for any finite $N$.
In this case their form, fixed up to a constant by conformal
invariance, is protected by a nonrenormalization theorem \cite{FGI}
valid for two- and three-point
functions of operators in the same multiplet as the stress tensor.

For any strong-weak coupling duality test it is essential to have
quantities that
do not acquire radiative corrections as one moves from weak to strong coupling.
If an exact
computation in the supergravity sector shows agreement with a
tree level result in the Yang-Mills sector, then there is an indication of a
nonrenormalization theorem at work.
This is the case for the three--point correlators
$<{\rm Tr} \Phi^{k_1}{\rm Tr} \Phi^{k_2}{\rm Tr} \Phi^{k_3}>$
computed in ref. \cite{MS} in the large $N$ limit of ${\cal N}=4$ $SU(N)$
Yang-Mills:
the  strong limit result $\lambda=g^2 N/4\pi \gg 1$ obtained using
type $IIB$ supergravity was shown to agree with the weak 't Hooft
coupling limit
$\lambda=g^2 N/4\pi \ll 1$ in terms of free fields. According to the AdS/CFT
correspondence one concludes that the correlators are independent of $\l$ to
leading order in $N$. A stronger conjecture made in ref. \cite{MS}
claims that three--point functions might be
independent of $g$ for {\em any} value of $N$.
As emphasized above, for the case $k=2$
nonrenormalization properties have been proven to be enjoyed by two--
and three--point functions of chiral operators.
For general $k$ there exists evidence of
nonrenormalization based on proofs that rely on reasonable assumptions
(analyticity
in harmonic superspace \cite{EHW} and validity of a generalized Adler-Bardeen
theorem \cite{PS}).

Explicit perturbative calculations in the
${\cal N}=4$  $SU(N)$ Yang-Mills conformal field theory are a way to confirm
the conjectures and add insights into potential larger symmetries of
the theory.
Important steps along this program have been made in \cite{HFS,S,HFMMR,BK}.
In particular it has been shown that to order
$g^2$ radiative
corrections do not affect the two- and three-point functions of chiral
operators \cite{HFS}.
The two-point function calculation has been performed for  chiral
operators with general $k$.
The result has been obtained showing that the order $g^2$ contribution
is proportional to the one for $k=2$ which indeed satisfies the
known nonrenormalization theorem mentioned above. The method cannot
be extended to higher
orders. In this paper  we address the non trivial test left open at
order $g^4$.
We find that corrections indeed vanish for {\em all} values of $N$, then
supporting the stronger conjecture of ref. \cite{MS}.

The paper is organized as follows:
in section 2 we briefly illustrate the ${\cal N}=4$  theory and
give the relevant rules for calculating
in ${\cal N}=1$ superspace. We present methods of performing
perturbation theory
calculations which make higher-loop correlators tractable. In Section 3
we test our approach computing the order $g^4$ contributions to the two-point
function with $k=2$ and check that their sum vanishes as it should. This  offers
us the opportunity to give details on the procedure, to collect results that
are useful for the subsequent calculation, to emphasize the points of interest
on which we will focus in the central part of the work. This is presented in
the following section where the two-point correlator with $k=3$ is
considered and the order $g^4$ contributions are collected.
The appendices contain a complete
list of our notations and details of the actual calculation.

\section{The general set up}

The physical particle content of ${\cal N}=4$ supersymmetric Yang-Mills
theory is given by one spin-$1$ Yang-Mills vector, four spin-$\frac{1}{2}$
Majorana spinors and six spin-$0$ particles in the ${\bf 6}$ of the
R--symmetry group $SU(4)$. All particles are massless and
transform under the adjoint representation of the $SU(N)$ gauge group.

Perturbative calculations are quite difficult to handle using a component field
formulation of the theory. (Note that in ref. \cite{HFS} a component
approach was used, but the order $g^2$ result for the two- and three-point
correlators was obtained using a
general argumentation based on colour combinatorics.
Only a schematic knowledge of the structure of the component action
was required.) In general, in order to resum Feynman diagrams at higher-loop
orders it is greatly advantageous to work in superspace.

In ${\cal N}=1$ superspace the action can be written in terms of one real
vector
superfield $V$ and three chiral superfields $\Phi^i$ containing the six
scalars organized into the ${\bf 3}\times \bf{ \bar 3}$ of $SU(3)
\subset SU(4)$ (we follow the notations in \cite{superspace})
\bea
S[J,\bar{J}]
&=&\int d^8z~ {\rm Tr}\left(e^{-gV} \Phib_i e^{gV} \Phi^i\right)+
\frac{1}{2g^2}
\int d^6z~ {\rm Tr} W^\a W_\a\nonumber\\
&&+\frac{ig}{3!} {\rm Tr} \int d^6z~ \e_{ijk} \Phi^i
[\Phi^j,\Phi^k]+\frac{ig}{3!} {\rm Tr} \int d^6\bar{z}~ \e_{ijk} \Phib^i
[\Phib^j,\Phib^k]\nonumber\\
&&+\int d^6z~ J {\cal O}+\int d^6\bar{z}~ \bar{J}\bar{{\cal O}}
\label{actionYM}
\eea
where $W_\a= i\Db^2(e^{-gV}D_\a e^{gV})$, and $V=V^aT^a$,
$\Phi_i=\Phi_i^a T^a$,
$T^a$ being $SU(N)$ matrices in the fundamental representation (see
Appendix A for a list of results on colour structures).
We have added to the classical
action source terms for the chiral primary operators generically denoted by
${\cal O}$ since our goal is the computation of their correlators.

Although in (\ref{actionYM}) the ${\cal N}=4$ supersymmetry invariance is
realized
only non linearly, the main advantage offered by a ${\cal N}=1$ formulation of
the theory resides in the fact that a straightforward off-shell quantum
formulation
is available. Thus if the aim is to perform higher-loop perturbative
calculations
this is the most suited approach to follow. Feynman rules are by now
standard and we have listed them in appendix B.

Now we focus on the two-point super-correlator for the operator
${\cal O}={\rm Tr}(\Phi^{\{ i} \Phi^j \Phi^{k\} })$.
As in ref. \cite{HFS}, we consider
the $SU(3)$ highest weight $\Phi^1$  field and compute
$<{\rm Tr}(\Phi^1)^3{\rm Tr}(\Phib^1)^3>$. This is not a restrictive
choice since all the other primary chiral correlators
can be obtained from this one by $SU(3)$ transformations.
What we gain is that we have no flavour
combinatorics and we are left to deal with the
colour combinatorics only.

We work in Euclidean space, with the generating functional defined as
\beq
W[J,\bar{J}]=\int {\cal D}\Phi~{\cal D}\Phib~{\cal D}V~e^{S[J,\bar{J}]}
\label{genfunc}
\eeq
Thus the two-point function is given by
\beq
<{\rm Tr}(\Phi^1)^3(z_1){\rm Tr}(\Phib^1)^3(z_2)>=
\left. \frac{\d^2 W}{\d J(z_1)\d\bar{J}(z_2)}\right|_{J=\bar{J}=0}
\label{defcorr}
\eeq
where $z \equiv (x,\theta, \bar{\theta})$.
We use perturbation theory to evaluate the contributions to $W[J,\bar{J}]$
which are
quadratic in the sources, i.e.
\beq
W[J,\bar{J}]\rightarrow \int d^4x_1~d^4x_2~ d^4\theta~
J(x_1,\theta,\bar{\theta})
\frac{F(g^2,N)}{(x_1-x_2)^6}\bar{J}(x_2,\theta,\bar{\theta})
\label{twopoint}
\eeq
where the $x$-dependence
of the result is fixed by the conformal invariance of the theory,
and $F(g^2,N)$ is the function that we want to determine up to order $g^4$. We
will find a result valid for any $N$.

In order to perform the calculation we have found it convenient to work in
momentum space,
using dimensional regularization and minimal subtraction scheme.
In $n$ dimensions, with $n=4-2\e$, the Fourier transform of the power factor
$(x_1-x_2)^{-6}$ in (\ref{twopoint}) is given by (see (\ref{Fourier}))
\beq
\frac{1}{(x^2)^3}=
\frac{\pi^{-2+\e}}{64} \frac{\G(-1-\e)}{\G(3)}
\int d^n p ~\frac{e^{-ipx}}{(p^2)^{-1-\e}}
\label{basicformula}
\eeq
The presence of the singular factor $\G(-1-\e) \sim
\frac{1}{\e}$ signals, in momentum space and in dimensional regularization,
the UV divergence of the correlation
function in (\ref{twopoint})
associated to the short--distance behaviour for $x_1 \sim x_2$.
It follows that performing perturbative calculations in momentum space
it is sufficient to look for all the contributions to (\ref{twopoint}) that
behave like $1/\e$, therefore disregarding finite contributions.
In fact, once the divergent terms are determined at a given
order in $g$, using
(\ref{basicformula}) one can reconstruct an $x$--space structure as in
(\ref{twopoint}) with a non--vanishing contribution to $F(g^2,N)$.
Finite contributions in momentum space
would correspond in $x$--space to terms proportional
to $\e$ which give rise only to contact terms \cite{CT}.

The one stated above is the basic rule of our strategy that
we can summarize as follows:
\begin{itemize}
\item consider all the two-point
diagrams from $W[J,\bar{J}]$ with $J$ and $\bar{J}$ on the external legs,
\item evaluate
all factors coming from combinatorics of the diagram and compute the colour
structure using the formulae collected in appendix A,
\item perform the
superspace $D$-algebra following standard techniques,
\item reduce the result to a multi-loop
momentum integral,
\item compute its $1/\e$ divergent contribution.
\end{itemize}
This last step, i.e. the
calculation of the divergent part of the various integrals we have achieved
using the method of uniqueness \cite{kazakov} and various rules and identities
\cite{CT,russians} that we have collected in appendix B. Since the theory is
at its conformal point, it is not affected by IR divergences. Therefore,
even if we work in a massless regularization scheme, we never worry about
the IR behavior of our integrals. Moreover, since the theory is finite,
the diagrams that we consider do not possess UV divergent subdiagrams.
Finally, as a general remark we observe that gauge-fixing the classical action
requires the
introduction of corresponding Yang-Mills ghosts. However they only
couple to the vector multiplet and do not enter our specific calculation.

In the next section, in order to get familiar with the formalism and to check
its validity,
we present the order $g^4$ calculation of the two-point correlator with $k=2$,
a non trivial test of our techniques.

\section{Preliminary test:$<{\rm Tr}(\Phi^1)^2{\rm Tr}(\Phib^1)^2>$ to order
$g^4$}

The two-point correlator we are interested in is obtained from $W[J,\bar{J}]$
inserting
in the action (\ref{actionYM}) the chiral operators ${\cal O}={\rm Tr}
(\Phi^1)^2$ and
$\bar{{\cal O}}={\rm Tr}(\Phib^1)^2$. As outlined in the previous section, the
relevant
contribution is obtained from the generating functional isolating terms of the
form
\beq
W[J,\bar{J}]\rightarrow \int d^4x_1~d^4x_2~ d^4\theta~ J(x_1,\theta,
\bar{\theta})
\frac{E(g^2,N)}{(x_1-x_2)^4}\bar{J}(x_2,\theta,\bar{\theta})
\label{twopointtwo}
\eeq
The general form of (\ref{twopointtwo}) is fixed by conformal invariance, while
the function $E(g^2,N)$ is the unknown to be determined. Fourier transforming
from $x$-space to momentum space
\beq
\frac{1}{(x^2)^2}=
\frac{\pi^{-2+\e}}{16} \frac{\G(-\e)}{\G(2)}
\int d^n p ~\frac{e^{-ipx}}{(p^2)^{-\e}}
\label{basicformulatwo}
\eeq
makes it clear that non--trivial contributions to the generating functional
are given by the divergent part of our Feynman diagrams.

\vskip 18pt
\noindent
\begin{minipage}{\textwidth}
\begin{center}
\includegraphics[width=0.40\textwidth]{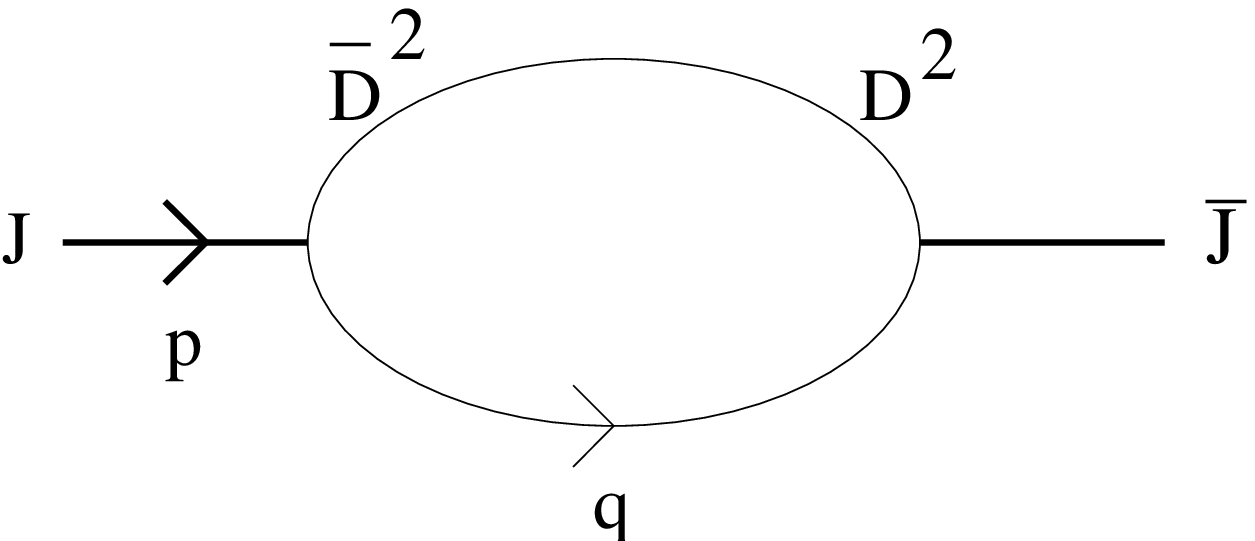}
\end{center}
\begin{center}
{\small{Figure 1:
tree--level contribution to $<{\rm Tr}(\Phi^1)^2 {\rm Tr}(\Phib^1)^2>$}}
\end{center}
\end{minipage}

\vskip 20pt
To start with we consider the tree-level contribution corresponding to the
graph in
Fig. 1. The $D$-algebra in this case is trivial and the one-loop momentum
integral gives
(see (\ref{1loop}))
\beq
\int \frac{d^n q}{q^2(p-q)^2}= \frac{\G(\e)\G^2(1-\e)}{\G(2-2\e)}
\frac{1}{(p^2)^\e}
\rightarrow \frac{1}{\e}
\label{oneloopint}
\eeq
Performing the trace operation on the colour indices and inserting factors
from combinatorics and Fourier transformations, we obtain
\beq
{\rm Fig}.~1~~~\rightarrow~~~ \frac{1}{(4\pi)^2}~2(N^2-1)~\frac{1}{\e}
\int d^4p~d^4\theta ~ J(-p,\theta,\bar{\theta})\bar{J}(p,\theta,\bar{\theta})
\label{treetwo}
\eeq

\vskip 18pt
\noindent
\begin{minipage}{\textwidth}
\begin{center}
\includegraphics[width=0.65\textwidth]{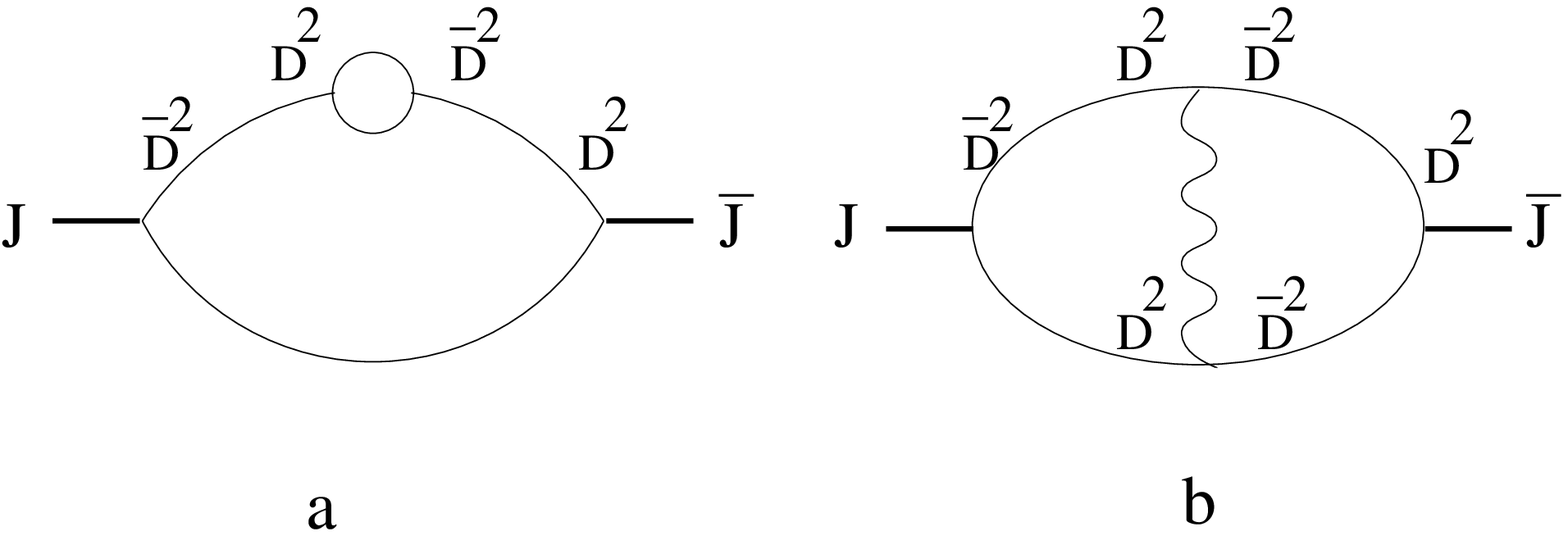}
\end{center}
\begin{center}
{\small{ Figure 2:
$g^2$--order contribution to $<{\rm Tr}(\Phi^1)^2 {\rm Tr}(\Phib^1)^2>$}}
\end{center}
\end{minipage}

\vskip 20pt

The order $g^2$ contribution, once evaluated in superspace gives
immediately a zero result:
the diagrams one would need to consider are shown in Fig. 2.
Diagram $2a$ does not contribute
since the one-loop correction to the chiral propagator vanishes due
to a complete cancellation
between vector and chiral loops \cite{GSR}. Diagram $2b$,
after completion of the $D$-algebra leads to a finite momentum integral
(see eq. (\ref{2loop})).

\vspace{0.8cm}

Now we consider the order $g^4$ contributions: they are shown in Fig. 3.
In Fig. $3a$ we have the insertion of a two-loop propagator
correction \cite{GSR}
\bea
&&-2g^4~N^2~k_2 ~\Phib^i_a(p,\theta,\bar{\theta})~ 
\Phi^i_a(-p,\theta,\bar{\theta})~p^2~
\int \frac{d^n q~d^nk}{k^2q^2(k-q)^2(k-p)^2(p-q)^2}\nonumber\\
&&~~~~~~~~~~~=-2g^4~N^2~k_2~ \Phib^i_a(p,\theta,\bar{\theta})~ 
\Phi^i_a(-p,\theta,\bar{\theta})~\frac{1}{(p^2)^{2\e}}
[6\zeta(3)+{\cal O}(\e)]
\label{2prop}
\eea
The two-loop integral has been evaluated using (\ref{2loop}).
\vskip 18pt
\noindent
\begin{minipage}{\textwidth}
\begin{center}
\includegraphics[width=1.00\textwidth]{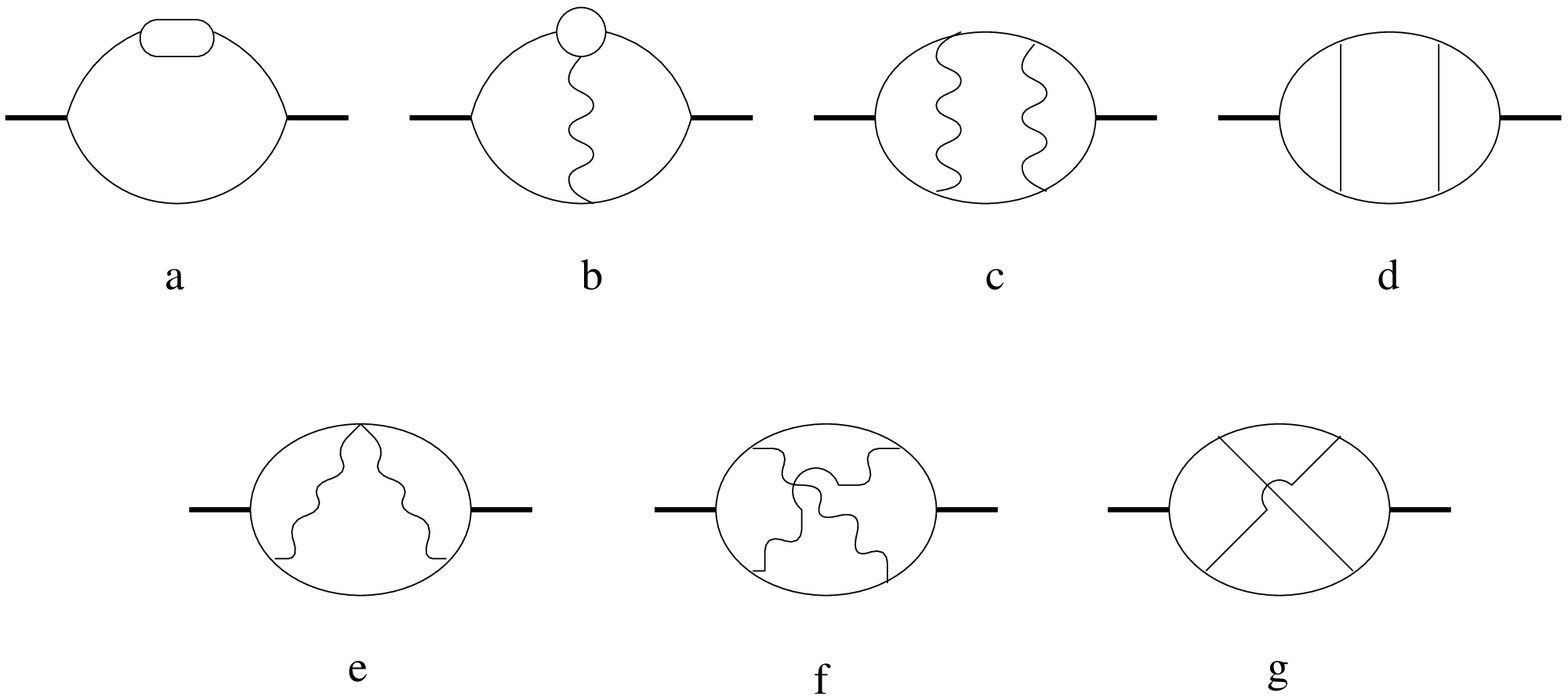}
\end{center}
\begin{center}
{\small{ Figure 3:
$g^4$--order  contribution to $<{\rm Tr}(\Phi^1)^2 {\rm Tr}(\Phib^1)^2>$}}
\end{center}
\end{minipage}

\vskip 20pt

In Fig. $3b$ a one-loop vertex correction appears: it corresponds
to the effective vertex \cite{GSR}
\bea
&&\frac{g^3}{4}~ N ~k_2~if^{abc} ~\Phib^i_a (q,\theta,\bar{\theta})
\Phi^i_b(-p,\theta,\bar{\theta})
\left( 4D^\a\Db^2D_\a \right. \nonumber\\
&&~~~~~~~~~~~~~~~~~~~~\left.+(p+q)^{\a\ad}
[D_\a,\Db_\ad ]\right) V_c (p-q,\theta,\bar{\theta})~
\int \frac{d^nk}{k^2(k-p)^2(k-q)^2}
\label{3vertex}
\eea
Note that a diagram with a vector propagator corrected at order $g^2$
is absent since at one--loop order there is a complete cancellation
among chiral, vector and ghost contributions \cite{GSR}.

A straightforward computation of the $D$-algebra for the diagrams in Fig.
$3c,3d,3e$ allows to
conclude that the corresponding momentum integrals are
actually all finite and, as previously observed, not relevant for our purpose.
More precisely we have for the diagrams in Fig. $3c, 3d$
\beq
p^4\int \frac{d^nk~d^nq~d^nr}{k^2q^2r^2(k-q)^2(q-r)^2(p-k)^2(p-q)^2(p-r)^2}
\label{3c3d}
\eeq
and for the diagram in Fig. $3e$
\beq
p^2\int\frac{d^nk~d^nq~d^nr}{k^2q^2(p-k)^2(p-r)^2(p-q)^2(k-r)^2(r-q)^2}
\label{3e}
\eeq
The above integrals are finite by power counting.
In general  we can disregard all the terms that in the course of the
$D$-algebra end up
with spinor derivatives $D$ or $\Db$ acting on the external legs: the
resulting momentum
integrals are finite and not interesting. Keeping this rule in mind
the evaluation of
the remaining diagrams is greatly simplified.

The graph in Fig. $3a$ is easy to compute. Using the result in (\ref{2prop})
and the one--loop integrals listed in Appendix B, with an overall factor
\beq
16\frac{1}{(4\pi)^6}g^4~N^2(N^2-1)~
\int d^4p~d^4\theta~J(-p,\theta,\bar{\theta})\bar{J}(p,\theta,\bar{\theta})
\label{overall}
\eeq
one obtains the following divergent contribution
\bea
&&{\rm Fig.~3a}~\rightarrow~3\z(3)\int\frac{d^nq}{(p-q)^2(q^2)^{1+2\e}}
\nonumber\\
&&~~~~~~~~~\rightarrow \zeta(3)\frac{1}{\e}
\label{3a}
\eea
From Fig. $3b$ we obtain a contribution only from
the $D^\a\Db^2D_\a $ term in the vertex (\ref{3vertex}), with the same
overall factor
 as in (\ref{overall})
\bea
&&{\rm Fig.~3b} ~\rightarrow~-
\int \frac{d^nk~d^nq~d^nr}{k^2(k-q)^2(k-r)^2q^2(q-r)^2}\frac{1}{(p-r)^2}
\nonumber\\
&&~~~~~~~~\rightarrow~-6\zeta(3)
\int \frac{d^nr}{(p-r)^2(r^2)^{1+2\e}}\nonumber\\
&&~~~~~~~~\rightarrow~-2\zeta(3)\frac{1}{\e}
\label{3b}
\eea
where we have used (\ref{2loop}) and (\ref{1loop}).

In the same way one analyzes the graph in Fig. $3f$: after completion of the
$D$-algebra one is left with a divergent integral as the one in (\ref{3b}).
Factoring out the same overall quantity we have
\beq
{\rm Fig.~3f}~\rightarrow~\frac{1}{2}\zeta(3)\frac{1}{\e}
\label{3f}
\eeq
Exactly the same result is obtained for the last diagram drawn in Fig. $3g$
\beq
{\rm Fig.~3g}~\rightarrow~\frac{1}{2}\zeta(3)\frac{1}{\e}
\label{3g}
\eeq
It is a trivial matter to sum up the contributions listed in (\ref{3a},
\ref{3b}, \ref{3f},
\ref{3g}) and obtain a vanishing result, as expected from the
nonrenormalization theorem.

Before closing this section we note that the diagrams in Figs. $3f$ and $3g$
lead to planar
contributions (i.e. with exactly the same $N$ dependence from colour
combinatorics as
the other diagrams): indeed to this order nonplanar diagrams are absent.
In the next section we will be confronted with a more complicated situation.

\section{The main calculation: $<{\rm Tr}(\Phi^1)^3{\rm Tr}(\Phib^1)^3>$
to order $g^4$}

Now we present the computation of the two-point function for the chiral
operator
${\cal O}={\rm Tr}(\Phi^1)^3$. To this end we go back to (\ref{twopoint})
and compute
the perturbative contributions to the function $F(g^2,N)$.
As previously emphasized,
making use of (\ref{basicformula}) we write Feynman diagrams in momentum
space and
isolate the $1/\e$ poles.

\vskip 18pt
\noindent
\begin{minipage}{\textwidth}
\begin{center}
\includegraphics[width=0.40\textwidth]{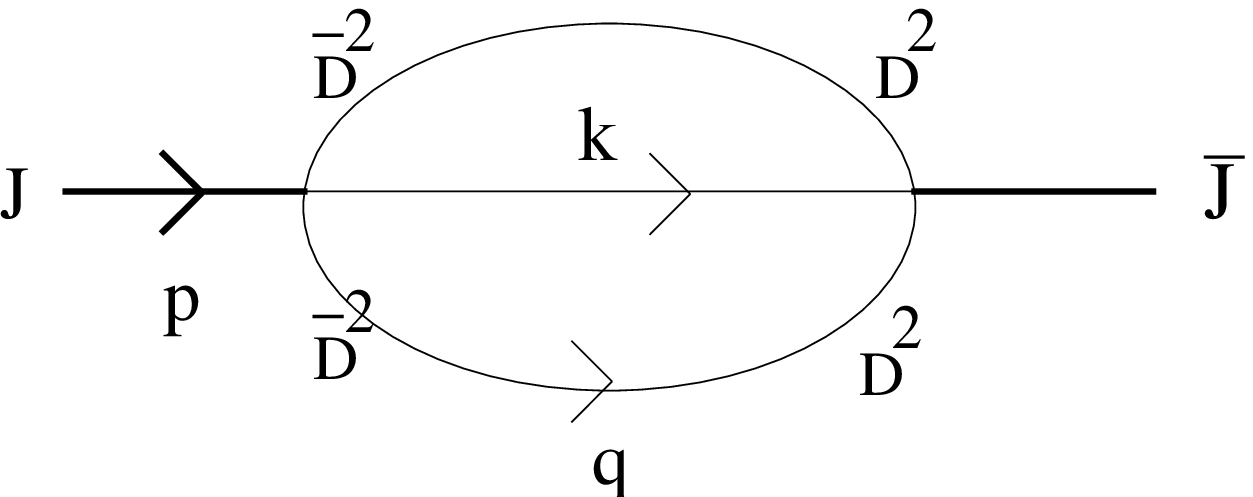}
\end{center}
\begin{center}
{\small{ Figure 4:
tree--level contribution to $<{\rm Tr}(\Phi^1)^3 {\rm Tr}(\Phib^1)^3>$}}
\end{center}
\end{minipage}

\vskip 20pt

In Fig. 4 we have drawn the tree-level contribution. The colour combinatorics
is evaluated with the
help of (\ref{treetraces}). With an overall factor
\beq
\frac{3}{(4\pi)^4}~\frac{(N^2-1)(N^2-4)}{N}
\int d^4p~d^4\theta~J(-p,\theta,\bar{\theta})\bar{J}(p,\theta,\bar{\theta})
\eeq
we obtain
\bea
&&{\rm Fig.~4}~\rightarrow~\int \frac{d^nq~d^nk}{q^2k^2(p-q-k)^2}\nonumber\\
&&~~~~~~~~\rightarrow -\frac{1}{4\e}p^2
\label{3tree}
\eea
The result in $x$-space is readily recovered using (\ref{basicformula}).

\vskip 18pt
\noindent
\begin{minipage}{\textwidth}
\begin{center}
\includegraphics[width=0.65\textwidth]{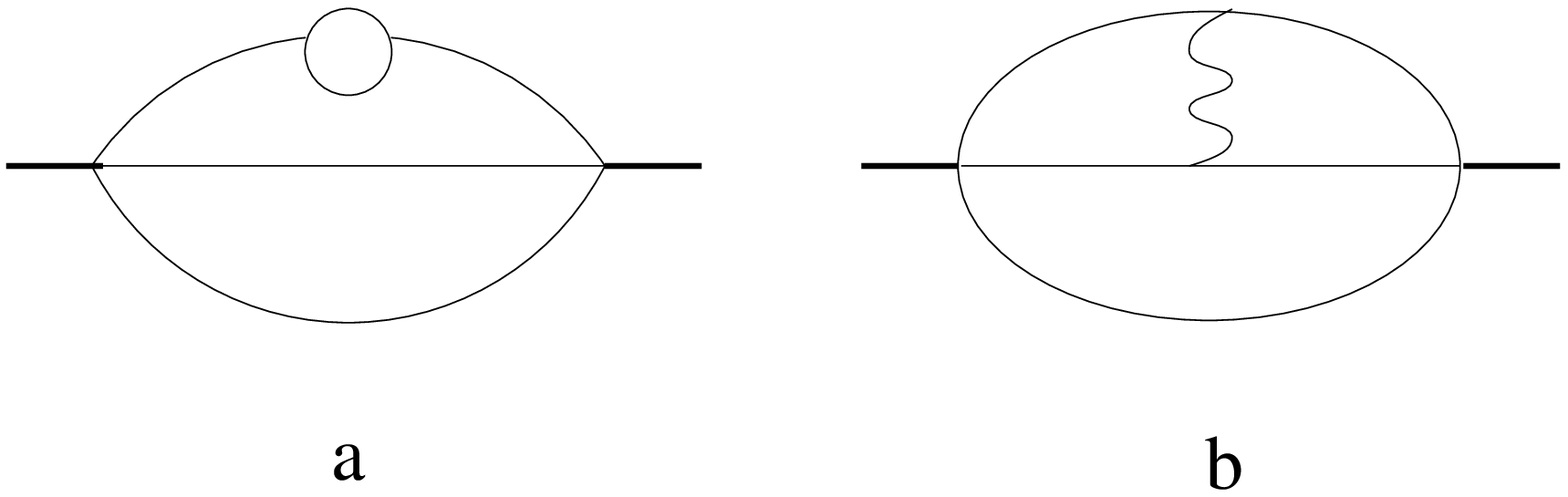}
\end{center}
\begin{center}
{\small{Figure 5:
$g^2$--order  contribution to $<{\rm Tr}(\Phi^1)^3 {\rm Tr}(\Phib^1)^3>$}}
\end{center}
\end{minipage}

\vskip 20pt

The superspace diagrams that enter the order $g^2$ computation are shown in
Fig. 5.
They are nothing else than the ones that appear in Fig. 2 with one line
added from
the chiral external vertices. One proves that their contributions vanish with
exactly
the same reasoning outlined in the previous section. As found in ref.
\cite{HFS} to order $g^2$ the vanishing
of the correlator is due to the fact that it is proportional to the correlator
of
${\cal O}={\rm Tr}(\Phi^1)^2$ for which the nonrenormalization theorem is
valid.
However, this is no longer true at order $g^4$ to which we turn now.

\vskip 18pt
\noindent
\begin{minipage}{\textwidth}
\begin{center}
\includegraphics[width=0.70\textwidth]{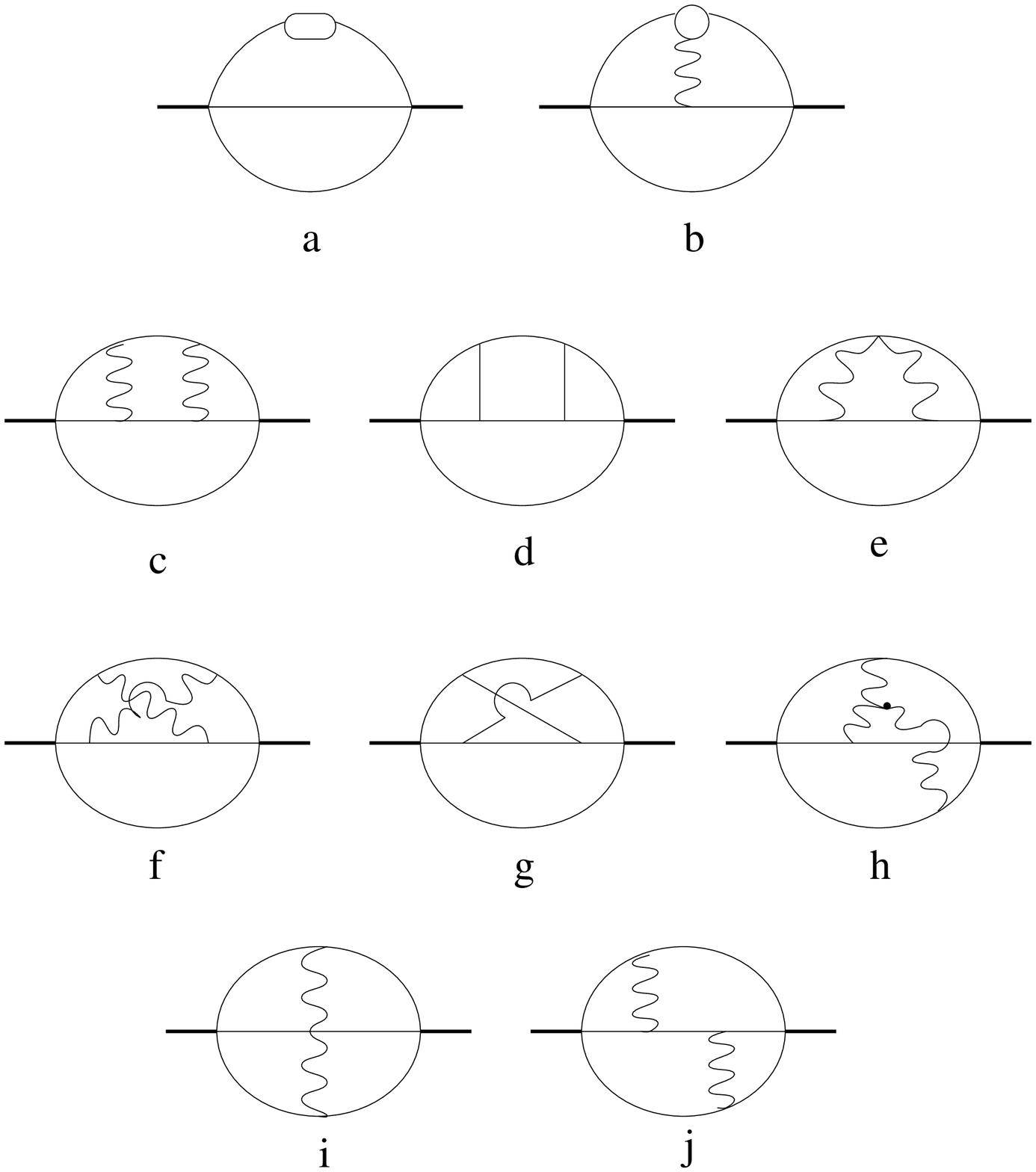}
\end{center}
\begin{center}
{\small{ Figure 6:
$g^4$--order  contribution to $<{\rm Tr}(\Phi^1)^3 {\rm Tr}(\Phib^1)^3>$}}
\end{center}
\end{minipage}

\vskip 20pt

The diagrams contributing to $g^4$--order are collected in Fig. 6.
The ones in Fig. $6a$--$6g$ are the same as in
Fig. 3 with one extra line added from the chiral external vertices. From the
result obtained
in the previous case at order $g^4$ (see Section 3), we would be tempted to
believe that
these diagrams still sum up to zero. However this would be a wrong conclusion.
In fact, what makes
things different is that in $6f$ and $6g$
the addition of the extra line changes completely the topology of the diagrams
which become really {\em  nonplanar}. As a consequence, 
their colour combinatorics changes and their
$N$-dependence is distinct from the remaining planar diagrams $6a$--
$6e$. More specifically, in this case it turns out
that the nonplanar diagrams $6f$ and $6g$  lead to a vanishing colour
combinatorics factor (the details of the calculation are given in Appendix A).

The evaluation of the colour coefficient for the other nonplanar diagram
in Fig. $6h$ reveals again a vanishing contribution (see Appendix A for
details).
The fact that the nonplanar diagrams do not contribute indicates that
the final answer is going to be valid for all values of $N$,
independently of any
large $N$ limit. In light of this result it becomes challenging to prove
the cancellation of nonplanar diagrams to all orders in the Yang-Mills
coupling.

Going back to Fig. 6, one easily convinces oneself that for the graphs in
Fig. $6c,6d,6e$ the same analysis as in the previous section applies.
In this case
the addition of the chiral
line simply adds a $D^2\Db^2$ factor which accounts for the $D$-algebra
of one added loop; performing the
$D$-algebra in the diagrams one is left with finite integrals. More precisely
we obtain for the diagrams in Fig. $6c$, $6d$
\beq
A\int d^ns\frac{s^4}{(p-s)^2(s^2)^{2+3\e}}
\label{6c6d}
\eeq
and for the diagram in Fig. $6e$
\beq
B\int d^ns\frac{s^2}{(p-s)^2(s^2)^{1+3\e}}
\label{6e}
\eeq
where $A$ and $B$ are finite constants, as a consequence of
the results in (\ref{3c3d}) and
(\ref{3e}). The final $s$-integration can be performed
using (\ref{1finite}) and a complete finite result is obtained in both
cases (\ref{6c6d})
and (\ref{6e}). Thus these terms are not relevant and we dismiss them.
We note that at this order diagrams containing the
scalar superpotential vertex
$\e_{ijk} {\rm Tr} (\Phi^{i} [\Phi^{j},\Phi^{k}])$ do not contribute.

We are left with the contributions from Fig. $6a,6b,6i,6j$ that we analyze
one by one. We will find that a highly nontrivial cancellation occurs.

For every diagram we need compute the specific combinatorics, the
various factors from vertices and propagators and the colour structure. Then
we have to perform the $D$-algebra in the loops  and finally evaluate the
momentum integrals. We factorize for each contribution the same quantity
\beq
\frac{9}{(4\pi)^8}~g^4 ~N(N^2-4)(N^2-1)\int d^4p~d^4\theta~
J(-p,\theta,\bar{\theta}) \bar{J}(p,\theta,\bar{\theta})
\label{relevantoverall}
\eeq
The diagram in Fig. $6a$ which contains the two-loop propagator correction in
(\ref{2prop}) is evaluated directly, using (\ref{1loop}) for the two successive
one-loop integrations,
\bea
&&{\rm Fig.}~6a~
\rightarrow~12\z(3)\int \frac{d^nk~d^nq}{k^2(q^2)^{1+2\e}(k+q-p)^2}
\nonumber\\
&&~~~~~~~~~~\rightarrow -\frac{3}{2}\z(3)\frac{1}{\e}~p^2
\label{6a}
\eea
The diagram in Fig. $6b$ contains the one-loop vertex correction given in
(\ref{3vertex}). In this case too, after completion of the $D$-algebra,
one obtains a relevant contribution only from the term in (\ref{3vertex})
which has the $D^\a\Db^2 D_\a$ factor. The resulting momentum integral is
\beq
{\rm Fig.}~6b~\rightarrow~ -4\int \frac{d^nk~d^nq~d^nr~d^nl}
{k^2(k-q)^2(r-q)^2(p-r)^2l^2(l-k)^2(l-q)^2}
\eeq
First using (\ref{2loop}) one performs the $k$ and $l$ integrations, then
with the help of (\ref{1loop}) one evaluates
the integrals on the $q$ and the $r$ variables
\beq
{\rm Fig.}~6b~\rightarrow~3\z(3)\frac{1}{\e}~ p^2
\label{6b}
\eeq
In the same way for the graph in Fig. $6i$ one has (see formula (\ref{4loop}))
\bea
&&{\rm Fig.}~6i~\rightarrow~-~p^2~\int\frac{d^nk~d^nq~d^nr~d^ns}{k^2q^2(k-q)^2
(p-k-r)^2(p-q-s)^2(r-s)^2r^2s^2}\nonumber\\
&&~~~~~~~~~\rightarrow~-5\z(5)\frac{1}{\e}~p^2
\label{6i}
\eea
Finally we concentrate on the evaluation of the diagram in Fig.~$6j$ which
after $D$-algebra gives rise to the following momentum integral
\beq
{\rm Fig.}~6j~\rightarrow~2\int
\frac{d^nk~d^nq~d^nr~d^ns~~~~~~~(p-r)^2(p-q)^2}{k^2(k-q)^2q^2r^2(r-s)^2
s^2(p-k-r)^2(p-q-r)^2(p-q-s)^2}
\eeq
The divergent contributions from this integral are obtained expanding
the numerator and keeping only terms with at the most $p^2$--powers.
Exploiting the symmetry of the integrand under $r \leftrightarrow q$ and
$k \leftrightarrow s$ we replace the numerator as 
\beq
(p-r)^2(p-q)^2~\rightarrow~2p^2r^2+r^2q^2-4r^2 p\cdot q +4p\cdot r ~p\cdot q
\eeq
We denote the integrals corresponding to the four terms by $J_1$, $J_2$,
$J_3$ and $J_4$ respectively and analyze them one by one.
Using (\ref{4tilde}) we obtain
\beq
J_1~\rightarrow~20\z(5)\frac{1}{\e}~p^2
\label{j1}
\eeq
Using (\ref{3tilde}) we obtain
\beq
J_2~\rightarrow~-\frac{3}{2}\z(3)\frac{1}{\e}~p^2
\label{j2}
\eeq
Simple manipulations which exploit the symmetries of the integrand
allow to reduce
the $J_3$ integral to the one in (\ref{4tilde}), so that we obtain
\beq
J_3~\rightarrow~-10\z(5)\frac{1}{\e}~p^2
\label{j3}
\eeq
We are left with the evaluation of $J_4$ which requires a more complicated
reasoning. The integral we are dealing with is
\beq
J_4= 8p^\m p^\n\int \frac{ d^nk~d^nq~d^nr~d^ns~~~~~~~~~r_\m q_\n}
{k^2(k-q)^2q^2r^2(r-s)^2s^2(p-k-r)^2(p-q-r)^2(p-q-s)^2}
\label{j4tutto}
\eeq
We observe that the divergent part of the integral is proportional to
$\d_{\m\n}$ so that restricting our attention to that part we can write
\beq
J_4^{div}~\rightarrow~\frac{8}{n}p^2\int~
\frac{d^nk~d^nq~d^nr~d^ns~~~~~~~~~r \cdot q}{k^2(k-q)^2q^2r^2(r-s)^2
s^2(p-k-r)^2(p-q-r)^2(p-q-s)^2}
\label{J4}
\eeq
To evaluate the divergent contribution in (\ref{J4}) we exploit the
result in eq. (\ref{4loop}) by first rewriting that integral as
\beq
I_4=\int~\frac{d^nk~d^nq~d^nr~d^ns~~~~~~~
(p-q-r)^2}{(p-q-r)^2
k^2q^2(k-q)^2(p-k-r)^2(p-q-s)^2(r-s)^2r^2s^2}
\eeq
where we have multiplied and divided by a factor $(p-q-r)^2$. Now we expand
the square in the numerator and neglect terms which, by power counting,
generate finite integrals. Using the result in (\ref{4tilde}) with a suitable
redefinition of the integration variables, we can write
\bea
I_4^{div} &=&
\int~\frac{d^nk~d^nq~d^nr~d^ns~~~~~~~q^2+r^2+2r\cdot q}{(p-q-r)^2k^2q^2
(k-q)^2(p-k-r)^2(p-q-s)^2(r-s)^2r^2s^2}
\nonumber \\
&=& 2~ \tilde{I}_4^{div} ~+~ \left( \frac{4}{n}p^2 \right)^{-1} J_4^{div}
\label{i4}
\eea
On the other hand, from eqs. (\ref{4loop}) and (\ref{4tilde}) we read
\beq
I_4^{div} ~=~ 5\z(5)~\frac{1}{\e} \qquad \qquad \tilde{I}_4^{div} =
5\z(5)~\frac{1}{\e}
\label{i4div}
\eeq
Comparing eqs. (\ref{i4}) and (\ref{i4div}) we obtain
\bea
&&J_4^{div}=\frac{4}{n} p^2 [ I_4^{ div}-2\tilde{I}_4^{div}]
\nonumber\\
&&~~~\rightarrow~-5\z(5)\frac{1}{\e}~p^2
\label{j4}
\eea
Finally summing the divergences from (\ref{j1}, \ref{j2}, \ref{j3}, \ref{j4})
we have
\beq
Fig.~6j~\rightarrow~[5\z(5)-\frac{3}{2}\z(3)]\frac{1}{\e}~p^2
\label{6j}
\eeq
At this point it is simple to add the four contributions in
(\ref{6a}, \ref{6b}, \ref{6i}, \ref{6j}) and check the complete cancellation
of the $1/\e$ terms. It is interesting to note that, while the diagrams 
$6a, ~6b$
only contribute with a divergent term proportional to $\z(3)$ and the diagram
$6i$ gives only a $\z(5)$--term, from the diagram $6j$
both terms arise with the correct coefficients to cancel completely the
divergence.

\section{Conclusions}

We have computed perturbatively up to $g^4$--order the two point
correlation function for the chiral primary operator ${\rm Tr} \Phi_1^3$
in ${\cal N} =4$ SU($N$) SYM theory. We have found a complete
cancellation of quantum corrections for any finite $N$.
Our result represents the first ${\cal O}(g^4)$ direct check of the
nonrenormalization theorem conjectured on the basis
of the AdS/CFT correspondence \cite{MS}. It supports also
the stronger claim \cite{MS} that there might be no quantum corrections at all,
for any finite $N$.

We have performed the calculation in ${\cal N}=1$ superspace using dimensional
regularization. The loop--integrals have been evaluated in momentum space
with the method of uniqueness \cite{kazakov, russians}.
In momentum space nontrivial, potential contributions appear as local divergent
terms that are easily isolated and evaluated. Finite contributions
would correspond to contact terms and can be neglected.

Our procedure is applicable to the perturbative analysis of more complicated
cases. Two--point functions for ${\rm Tr} \Phi^k$, $k>3$,
three--point functions and extremal correlators for chiral primary
operators are now under investigation.

\medskip

\section*{Acknowledgements}
\noindent We wish to thank Dan Freedman for his continuous encouragement,
enthusiasm and
helpful suggestions during the course of this calculation.\\
\noindent This work has been supported by the European Commission TMR
programme ERBFMRX-CT96-0045, in which S.P. and D.Z. are associated
to the University of Torino.

\newpage

\appendix

\section{Colour conventions and relevant identities}

In this Appendix we give our conventions and list a set of colour structure
identities that we have used
in the course of the calculation. In addition we evaluate explicitly
the colour coefficients for the $g^4$--order nonplanar diagrams in
Figs. $6f$, $6g$ and $6h$.

The $SU(N)$ generators $\{T_a\}$, $a=1, \cdots, N^2-1$, satisfy the algebra
\beq
[T_a, T_b] ~=~ if_{abc} T_c
\eeq
where $f_{abc}$ are the structure constants.
In the fundamental representation they are given by $N \times N$ traceless
matrices satisfying
\beq
{\rm Tr} (T_a T_b)= k_2 ~\d_{ab}
\eeq
Here $k_2$ is an arbitrary constant (usually $k_2 = \frac12$) which can
be eliminated by a suitable rescaling of the $T_a$--matrices and the structure
constants. We leave it unspecified, since in the evaluation of
correlation functions it only appears at an intermediate stage, the
final answer being independent of $k_2$.

The anticommutator of two generators defines a totally symmetric tensor
$d_{abc}$, according to the relation
\beq
\{T_a,T_b\} =2k_2 \left( \frac{1}{N} \d_{ab}+d_{abc} T_c\right)
\eeq
Now, using the previous definitions and the relation
\beq
T^a_{ij} T^a_{kl}= k_2 \left( \d_{il}\d_{jk}-
\frac{1}{N}\d_{ij}\d_{kl}\right)
\eeq
one can easily derive the following identities
\beq
(T^aT^bT^a)_{ij}= -\frac{k_2}{N} T^b_{ij}
\label{a4}
\eeq
\beq
{\rm Tr}(T_aT_bT_c)=\frac{k_2}{2}[(2k_2)d_{abc}+if_{abc}]
\label{a8}
\eeq
\beq
f_{acd}f_{bcd}= 2k_2 N\d_{ab} \qquad\qquad d_{acd}d_{bcd}= \frac{1}{2k_2}
\frac{N^2-4}{N}\d_{ab}
\label{a1}
\eeq
\beq
f_{alm}f_{bmn}f_{cnl}=k_2Nf_{abc}
\label{a2}
\eeq
\beq
d_{abc}d_{aem}d_{bdm}=\frac{1}{k_2}\frac{N^2-12}{4N}d_{cde}
\label{a3}
\eeq
\beq
{\rm Tr}(T_aT_bT_c)d_{abd}=k_2\frac{N^2-4}{2N}\d_{cd}
\label{a5}
\eeq
\beq
{\rm Tr}(T_aT_bT_c)f_{amd}f_{bme}=k_2 N{\rm Tr}(T_cT_dT_e)
\label{a7}
\eeq
\beq
{\rm Tr}(T_aT_bT_c)d_{amd}d_{bme}=
\frac{1}{k_2}\left[\frac{N^2-8}{4N}{\rm Tr}(T_dT_eT_c)-\frac{1}{N}{\rm Tr}
(T_eT_dT_c)\right]
\label{a6}
\eeq
\beq
{\rm Tr}(T_aT_bT_c)[{\rm Tr}(T_aT_bT_c)+{\rm Tr}(T_aT_cT_b)]=k_2^3\frac{(N^2-4)(N^2-1)}{N}
\label{treetraces}
\eeq
For the product of four structure constants one obtains
\bea
&& f_{amn} f_{bnp} f_{cpr} f_{drm} = \nonumber \\
&~& ~~~~~~~~~~~(2k_2)^2 \left[ \d_{ab} \d_{cd} +
\d_{ad} \d_{bc} + k_2 \frac{N}{2} ( d_{abe} d_{cde} - d_{ace} d_{bde}
+ d_{ade} d_{bce}) \right]
\label{a9}
\eea
Finally, from Jacobi identity one derives
\beq
f_{abm}f_{cdm} + f_{cbm}f_{dam} + f_{dbm}f_{acm} = 0
\label{jacobi}
\eeq
\vspace{0.8cm}

Now we concentrate on the evaluation of the colour coefficients for
the diagrams $6f$ and $6g$. The colour structure
arising from these diagrams is
\beq
{\cal C} \equiv [{\rm Tr}(T_aT_bT_c)+{\rm Tr}(T_aT_cT_b)]
~f_{amn} f_{enp} f_{bpr} f_{drm} ~
[{\rm Tr}(T_dT_eT_c)+{\rm Tr}(T_dT_cT_e)]
\label{a10}
\eeq
Using the identities (\ref{a8}) and (\ref{a9}) the previous expression can
be rewritten as
\beq
{\cal C} ~=~ 16 k_2^6 \left[ 2 d_{abc} d_{abc} ~+~ k_2 \frac{N}{2}
\left( 2 d_{abc} d_{aem} d_{bdm} d_{dec} ~-~
d_{abc} d_{abm} d_{edm} d_{dec} \right) \right]
\eeq
We exploit the identities (\ref{a1}) and (\ref{a3}) for the product
of two and three $d$--tensors respectively, so that we obtain
\bea
{\cal C} &=& 8 k_2^5 \left\{ \frac{2}{N} (N^2-1)(N^2-4) ~+~
\frac{1}{4N} (N^2-1)(N^2-4) \left[ (N^2 -12) ~-~ (N^2-4) \right]
\right\} \nonumber \\
&=& 0
\ena
Finally, we prove the vanishing of the colour coefficient for the
nonplanar diagram $6h$.
The colour structure for this diagram is
\beq
[{\rm Tr}(T_aT_bT_c)+{\rm Tr}(T_aT_cT_b)]
~f_{amd} f_{bne} f_{cpg} f_{mnp} ~
[{\rm Tr}(T_dT_eT_g)+{\rm Tr}(T_dT_gT_e)]
\label{a11}
\eeq
This expression is trivially zero since the product of the four
structure constants is antisymmetric under the exchange $a \leftrightarrow
b$ and $d \leftrightarrow e$, whereas the rest of the expression
is symmetric.
However, a stronger result holds which states the vanishing of (\ref{a11})
without exploiting any symmetry of the expression. 
Let us concentrate, for example, on the first piece  
\beq
{\rm Tr}(T_aT_bT_c) f_{amd} f_{bne} f_{cpg} f_{mnp} ~ {\rm Tr}(T_dT_eT_g).
\eeq
By means of the Jacobi identity (\ref{jacobi}) applied to the product 
$f_{cpg} f_{mnp}$, this expression can be easily rewritten as 
\bea
&&{\rm Tr}(T_aT_bT_c)~[-f_{bne} f_{cnp} f_{dma} f_{gmp} + 
f_{amd} f_{cmp} f_{enb} f_{gnp}]~{\rm Tr}(T_dT_eT_g) \nonumber\\
&&~~~~~~~~~~~~~~~~~\nonumber \\
&& = k_2^2 N^2 [-{\rm Tr}(T_aT_eT_p){\rm Tr}(T_aT_eT_p) + 
{\rm Tr}(T_dT_bT_p){\rm Tr}(T_dT_bT_p)] = 0
\ena
where the identity (\ref{a7}) has been used. Analogous cancellations occur 
for the other three pieces of the sum in (\ref{a11}).

\section{Conventions and details of the calculation in momentum space}

The quantization procedure of the classical
action in (\ref{actionYM}) requires the introduction of a gauge fixing
and corresponding ghost terms.
We have found it convenient to work in Feynman gauge,
so that the vector  and the chiral superfield propagators are treated on the
equal footing. Here we do not repeat the various steps that lead to the
construction
of the quantum action (see for example \cite{superspace}); we simply give
the essential ingredients.
The ghost superfields only couple to the vector
multiplet and are not interesting for our calculation. The relevant interactions
are the ones in (\ref{actionYM}).
In momentum space we have the superfield propagators
\beq
<V^a V^b>= -\frac{1}{k_2} \frac{\d^{ab}}{p^2}\qquad\qquad
<\Phi^a_i \Phib^b_j>=\frac{1}{k_2}\d_{ij} \frac{\d^{ab}}{p^2}
\label{propagators}
\eeq
The vertices are read directly from the interaction terms in (\ref{actionYM}),
with additional $\Db^2$, $D^2$ factors for chiral, antichiral lines
respectively.
The ones that we need are the following
\bea
&&V_1=ig k_2 f_{abc}\d^{ij} \Phib^a_i V^b \Phi^c_j \qquad\qquad \qquad
V_2=-\frac{i}{2}gk_2 f_{abc}V^a \Db^2 D^\a V^b D_\a V^c \nonumber\\
&&~~~~~~~~~~~~~~~~~
V_3=\frac{g^2}{2} k_2  \d^{ij} f_{adm} f_{bcm} V^aV^b \Phib^c_i
\Phi^d_j \\
&& V_4 = - \frac{g}{3!} \e^{ijk} f_{abc} \Phi_i^a \Phi_j^b \Phi_k^c
\qquad \qquad \qquad
\bar{V}_4 = - \frac{g}{3!} \e^{ijk} f_{abc} \Phib_i^a \Phib_j^b \Phib_k^c
\nonumber
\label{vertices}
\eea
All the calculations are performed in $n$ dimensions with $n=4-2\e$.

We make use of the $n$-dimensional Fourier transform
\beq
\int d^np~\frac{e^{-ipx}}{(p^2)^\n}= \pi^{\frac{n}{2}} 2^{n-2\n}
\frac{\G(\frac{n}{2}-\nu)}
{\G(\n)} \frac{1}{(x^2)^{\frac{n}{2}-\n}} \qquad \quad
\nu \neq \frac{n}{2}, \frac{n}{2} +1, \cdots
\label{Fourier}
\eeq
and perform the calculation in momentum space.
The various integrals we have to deal with are all computed making use
of the method of uniqueness \cite{kazakov}. This approach, which
consists in
an analytical calculation of multiloop Feynman diagrams, is particularly
efficient
for massless Feynman integrals of a single variable (either momentum
or coordinate).
Since we are evaluating a two--point correlator
with fields that
are all massless, we are just there where the method best applies.

At intermediate stages of the calculation we drop $2\p$ factors and
reinstate them at the
end, multiplying the final result by a $1/(4\pi)^2$ factor for every loop.
With this understanding we give some of the integrals that we made use of:\\
At one loop
\beq
I_1=\int  \frac{d^nk}{(k^2)^\a [(p-k)^2]^\b}=
\frac{\G(\a+\b-\frac{n}{2} )}{\G(\a)\G(\b)} 
\frac{\G(\frac{n}{2}-\a)\G(\frac{n}{2} -\b)}
{\G(n-\a-\b)} \frac{1}{(p^2)^{\a+\b-\frac{n}{2} }}
\label{1loop}
\eeq
At two loops
\beq
I_2=\int \frac{d^nk~d^nq}{k^2q^2(k-q)^2(k-p)^2(q-p)^2}= \frac{1}{(p^2)^{1+2\e}}
[6\z(3)+{\cal O}(\e)]
\label{2loop}
\eeq
At three loops
\beq
I_3=\int \frac{d^nk~d^nq~d^nr}{q^2k^2(p-q)^2(r-q)^2(p-r)^2(r-k)^2(k-q)^2}=
\frac{1}{(p^2)^{1+3\e}}[20\z(5)+{\cal O}(\e)]
\label{3loop}
\eeq
At four loops
\beq
I_4=\int \frac{d^nk~d^nq~d^nr~d^ns}{k^2q^2(k-q)^2(p-k-r)^2(p-q-s)^2(r-s)^2
r^2s^2}
=\frac{1}{(p^2)^{4\e}}~\frac{1}{\e}~[5\z(5)+{\cal O}(\e)]
\label{4loop}
\eeq

\vspace{0.5cm}
From the previous integrals we can derive several other results that
we used in the course of our calculation:\\
from (\ref{1loop}) one obtains
\beq
\tilde{I}_0= \int \frac{d^nk}{(k^2)^{\a\e}(p-k)^2}=
-\frac{\a}{2(1+\a)}p^2+{\cal O}(\e)
\label{1finite}
\eeq
From (\ref{2loop}) we have
\bea
&&\tilde{I}_3=\int \frac{d^nk~d^nq~d^ns}{k^2q^2(k-q)^2(k-s)^2
(q-s)^2(p-s)^2}\nonumber\\
&&~~~~~~~~=[ 6\z(3)+{\cal O}(\e)]\int\frac{d^ns}{(s^2)^{1+2\e}(p-s)^2}
\nonumber\\
&&~~~~~~~~=\frac{1}{(p^2)^{3\e}}\frac{1}{\e}[2\z(3)+{\cal O}(\e)]
\label{3tilde}
\eea
In the same way from (\ref{3loop}) we have
\bea
&&\tilde{I}_4=\int \frac{d^nk~d^nq~d^nr~d^ns}{q^2k^2(s-q)^2(r-q)^2(s-r)^2
(r-k)^2
(k-q)^2(p-s)^2}\nonumber\\
&&~~~~~~~~=[20\z(5)+{\cal O}(\e)]\int \frac{d^ns}{(s^2)^{1+3\e}(p-s)^2}
\nonumber\\
&&~~~~~~~~=\frac{1}{(p^2)^{4\e}} \frac{1}{\e}[5\z(5)+{\cal O}(\e)]
\label{4tilde}
\eea


\newpage
\begin{thebibliography}{99}

\bibitem{adscft}J. Maldacena, Adv. Theor. Math. Phys. {\bf 2} (1998) 231,
hep-th/9711200;\\
S.S. Gubser, I.R. Klebanov and A.M. Polyakov, Phys. Lett. {\bf B428} (1998)
105, hep-th/9802109;\\
E. Witten, Adv. Theor. Math.  Phys. {\bf 2} (1998) 253, hep-th/9802150
\bibitem{FGI} S.S. Gubser and I. Klebanov, Phys. Lett. {\bf B413} (1997) 41,
hep-th/9708005;\\
D. Anselmi, D. Freedman, M. Grisaru and A. Johansen, Nucl. Phys. {\bf B526}
(1998) 543, hep-th/9708042.
\bibitem{MS} S. Lee, S. Minwalla, M. Rangamani and N. Seiberg, Adv. Theor.
Math. Phys. {\bf 2} (1998) 697, hep-th/9806074.
\bibitem{EHW}B. Eden, P.S. Howe and P.C. West, "Nilpotent invariants in
${\cal N}=4$ SYM", hep-th/9905085;\\
P.S. Howe, C. Schubert, E. Sokatchev and P.C. West,
"Explicit construction of nilpotent covariants in ${\cal N}=4$ SYM",
hep-th/9910011.
\bibitem{PS} A. Petkou and K. Skenderis, "A nonrenormalization theorem
for conformal anomalies", hep-th/9906030.
\bibitem{HFS} E. D'Hoker, D.Z. Freedman and W. Skiba, Phys. Rev. {\bf D59}
(1999) 45008, hep-th/9807098.
\bibitem{S} W. Skiba, "Correlators of short multitrace operators in
${\cal N}=4$ supersymmetric Yang-Mills", hep-th/9907088.
\bibitem{HFMMR} E. D'Hoker, D.Z. Freedman, S.D. Mathur, A. Matusis and
L. Rastelli,
"Extremal correlators in the AdS/CFT correspondence", hep-th/9908160.
\bibitem{BK} M. Bianchi and S. Kovacs, "Non-renormalisation of extremal
correlators in ${\cal N}=4$ SYM theory", hep-th/9910016.
\bibitem{superspace} S.J. Gates, M.T. Grisaru, M. Ro\v{c}ek and W. Siegel,
"Superspace" (Benjamin-Cummings, Reading, MA, 1983).
\bibitem{CT} K.G. Chetyrkin and F.V. Tkachov, Nucl. Phys.
{\bf B192} (1981) 159.
\bibitem{kazakov} D.I. Kazakov, Phys. Lett. {\bf 133B} (1983) 406;
Theor. Math. Phys. {\bf 43} (1985) 462.
\bibitem{russians} K.G. Chetyrkin, A.L. Kataev and F.V. Tkachov, Nucl. Phys.
{\bf B174} (1980) 345.
\bibitem{GSR} M.T. Grisaru, W. Siegel and M. Ro\v{c}ek, Nucl. Phys. {\bf B159}
(1979) 429.


\end{thebibliography}
\end{document}